\documentclass[aps,floatfix,nofootinbib]{revtex4}
\usepackage{graphicx}

\begin{document}

\title{An explicit global covering of the Schwarzschild-Tangherlini black holes}
\author{Kayll Lake \cite{email}}
\affiliation{Department of Physics and Department of Mathematics
and Statistics, Queen's University, Kingston, Ontario, Canada, K7L
3N6 }
\date{\today}

\begin{abstract}
Explicit regular coordinates that cover all of the Tangherlini
solutions (Schwarzschild black holes of dimension $D>4$) are
given. The coordinates reduce to Israel coordinates for $D=4$.
\end{abstract}
\maketitle
\bigskip
One consequence of string theory is the importance now attached to
the study of higher dimensional black hole backgrounds. Even at
five dimensions asymptotically flat stationary vacuum black holes
are not unique in the sense that horizons of topology $S^1 \times
S^2$ \cite{er} in addition to $S^3$ \cite{mp} are now known.
However, in the simpler static case, it is known that
asymptotically flat static vacuum black holes are unique
\cite{flat} and given by the Tangherlini generalization of the
Schwarzschild vacuum \cite{tangherlini}. The purpose of this note
is to exhibit global, regular and explicit coordinates for these
spaces. In addition to their obvious utility, these coordinates
may be useful for studies of stability in the nonlinear regime
\cite{stability}.

\bigskip
For dimensions $D\geq4$ the space
\begin{equation}
ds^2=-2\,{\frac {{w}^{2} ( - ( r) ^{-D+3}{2}^{D}{m}^{D}+ 4 m^2(D(2
m -r)+3r-4m ) }{m ( 2\,m-r) ^{2}}} dU^2+8 m dUdw+
r^2d\Omega_{D-2}^{2}, \label{metric}
\end{equation}
where $r \geq 0$ signifies the function
\begin{equation}
r( U,w ) \equiv 2\,m+( D-3) Uw, \label{rfn}
\end{equation}
$m$ is a constant $>0$ and $d\Omega_{D-2}^{2}$ is the metric of a
unit $D-2$ sphere \cite{plane}, satisfies
\begin{equation}
R_{a b}=0 \label{vacuum}
\end{equation}
where $R_{a b}$ is the Ricci tensor \cite{dimensions}. The
trajectories with $D$-tangents $k^{a}=\delta^{a}_{w}$ are null
geodesics affinely parameterized by $w$ and give
$r\nabla_{a}k^{a}=(D-2)(D-3)U$. For (\ref{metric}) there is but
one independent invariant derivable from the Riemmann tensor
without differentiation. This invariant can be taken to be the
square of the Weyl tensor $C_{a b c d}$ and for (\ref{metric})
\begin{equation}
C_{a b c d}C^{a b c d} \propto \frac{m^{2(D-3)}}{r^{2(D-1)}}
\label{weylsq}
\end{equation}
and so (\ref{metric}) is singular only at $r=0$ \cite{constant}.
The metric (\ref{metric}) is regular for $r>0$ although this is
manifest only on specification of $D$ due to the compact form
given. In particular, note that
\begin{equation}
g_{_{U U}}|_{_{r=2m}}=2(D-2)(D-3)w^2 \label{regular}.
\end{equation}
Explicit forms of $g_{_{U U}}$ follow from (\ref{metric}) and for
the cases $D=4...11$ they are given in Appendix A. With $U$ and
$w$ extending over the reals, (\ref{metric}) is geodesically
complete. For the case $D=4$ the coordinates are due to Israel
\cite{israel}.

\bigskip
For $D=4$ a regular covering of the vacuum manifold is usually
given in terms of the Kruskal-Szekeres coordinates (say $u$ and
$v$) \cite{kruskal} which, as is well known, are only implicit
since $r(uv)$ is a non-invertible function. The generalization of
these coordinates to $D$ dimensions has been considered by Gregory
and Laflamme \cite{gregory}. (A. Krasi\'nski has kindly pointed
out an earlier study \cite{krasinski}.) Now $r$ is defined by way
of a generalized tortoise coordinate so that
\begin{equation}
\pm
2muv=exp(\frac{D-3}{2m}\int\frac{r^{D-3}dr}{r^{D-3}-(2m)^{D-3}})
\equiv f(r)\frac{r-2m}{r+2m} \label{tortoise}.
\end{equation}
Relatively simple forms for $f(r)$ are given in Appendix B for
$D=4...11$ excepting the cases $D=8, 10$ for which no simple forms
where found. Clearly the explicit form (\ref{rfn}) is simpler. The
disadvantage of the coordinates $U,w$ lies in the fact that the
``other" radial null geodesics are given in terms of the
differential equation $dU/dw=-8m/g_{_{U U}}$. Compared to the
implicit nature of the Kruskal-Szekeres coordinates this
disadvantage might be considered minor in the sense that these
geodesics can always be constructed numerically. (Explicit
solutions can be found depending on $D$.) Note that for all $D$,
$dU/dw \rightarrow 0$ as $r \rightarrow 0$ and $dw/dU \rightarrow
0$ as $w \rightarrow 0$. The trajectories with $D$-tangents
$l^{a}=\delta^{a}_{U}$ have $r\nabla_{a}l^{a}=(D-2)(D-3)w$ and
become null geodesics at $w=0$ where they generate the other
branch to the horizon $r=2m$.
\begin{acknowledgments}
This work was motivated by lectures at Black Holes IV and
supported by a grant from the Natural Sciences and Engineering
Research Council of Canada. Portions of this work were made
possible by use of \textit{GRTensorII} \cite{grt}.
\end{acknowledgments}
\appendix
\section{Forms of $g_{_{U U}}$ for $D=4...11$}
 D=4 (Israel coordinates)
\begin{equation}
8\,{\frac {m{w}^{2}}{r}}
\end{equation}
D=5
\begin{equation}
16\,{\frac {{w}^{2} ( m+r  ) m}{ r^{2}}}
\end{equation}
D=6
\begin{equation}
8\,{\frac {{w}^{2} ( 4\,{m}^{2}+4\,mr+3\,
 r ^{2} ) m}{r
^{3}}}
\end{equation}
D=7
\begin{equation}
16\,{\frac {{w}^{2} ( 4\,{m}^{3}+4\,r{m}^{2}+ 3\,m r ^{2}+2\,
r^{3} ) m}{ r^{4}}}
\end{equation}
D=8
\begin{equation}
8\,{\frac {{w}^{2} ( 16\,{m}^{4}+16\,{m}^{3}r+12\, r
^{2}{m}^{2}+8\,m r ^{3}+5\, r  ^{4} ) m}{ r^{5}}}
\end{equation}
D=9
\begin{equation}
16\,{\frac {{w}^{2} ( 16\,{m}^{5}+16\,r{m}^{4 }+12\,{m}^{3} r
^{2}+8\, r ^{3}{m}^{2}+5\,m  r^{4}+3\,r^{5}
 ) m}{r ^{6}}}
\end{equation}
D=10
\begin{equation}
8\,{\frac {{w}^{2} ( 64\,{m}^{6}+64\,r {m}^{5} +48\, r
^{2}{m}^{4}+32\,{m}^{3}
 r ^{3}+20\, r ^{4}{m}^{2}+12\,m r^{5}+7\, r^{6} ) m}
{r^{7}}}
\end{equation}
D=11
\begin{equation}
16\,{\frac {{w}^{2} ( 64\,{m}^{7}+64\,r{m}^{6 }+48\, r
^{2}{m}^{5}+32\, r ^{3}{m}^{4}+20\,{m}^{3} r ^{4}+12\, r^{5
}{m}^{2}+7\,m r^{6}+4\, r ^{7} ) m}{ r ^{8}}}
\end{equation}
\section{Forms for $\pm 2m uv = f(r)\frac{r-2m}{r+2m}$  }
 D=4 (Kruskal-Szekeres coordinates)
\begin{equation}
( r+2\,m ) {e^{{\frac {r}{2m}}}}
\end{equation}
D=5
\begin{equation}
{e^{{\frac {r}{m}}}}
\end{equation}
D=6
\begin{equation}
{\frac { (r+2\,m) {e^{( \frac{3\,r}{2m}-\sqrt {3}\arctan
 ( {\frac {( r+m ) }{\sqrt {3}m}})
 )}}}{\sqrt {{r}^{2}+2\,rm+4\,{m}^{2}}}}
\end{equation}
D=7
\begin{equation}
{e^{2\, ( \frac{r}{m}-\arctan ({\frac {r}{2m}} )
 ) }}
\end{equation}
D=9
\begin{equation}
{e^{( \frac{3\,r}{m}-\sqrt {3}\arctan
 ( {\frac {( r+m ) }{\sqrt {3}m}})-\sqrt {3}\arctan
 ( {\frac {( r-m ) }{\sqrt {3}m}})
 )}}( {\frac {{r}^{2}-2\,mr\sqrt {2}+4\,{m}^{2}}{ {r}^{2}+2\,mr\sqrt
{2}+4\,{m}^{2}}} ) ^{\frac{1}{2}}
\end{equation}
D=11
\begin{equation}
{e^{ ( \frac{4r}{m}-2\arctan ({\frac {r}{2m}} ) - \sqrt {2}\arctan
({\frac {\sqrt {2}r-2\,m}{2m}} )- \sqrt {2}\arctan ({\frac {\sqrt
{2}r+2\,m}{2m}} )
 )}} ( {\frac {{r}^{2}-2\,mr\sqrt {2}+4\,{m}^{2}}{
{r}^{2}+2\,mr\sqrt {2}+4\,{m}^{2}}} ) ^{\frac{1}{\sqrt {2}}}
\end{equation}


\begin{thebibliography}{}
\bibitem{plane}The corresponding plane and hyperbollic cases are
handled similarly.
\bibitem{dimensions}We use geometrical units. Note that the
tensors here are $D\geq4$ dimensional.
\bibitem{constant} Under a conformal transformation $\widetilde{g}_{a
b}=\Omega^2 g_{a b}$, $\Omega=constant$, $\widetilde{R}_{a b}=R_{a
b}$ but $\Omega^4\widetilde{C}_{a b c d}\widetilde{C}^{a b c
d}=C_{a b c d}C^{a b c d}$ and so the constant of proportionality
in (\ref{weylsq}) is of no interest.
\end{thebibliography}
\end{document}